\documentclass[a4Apaper,11pt]{article}
\usepackage{jinstpub} 
\usepackage[percent]{overpic}
\usepackage{siunitx}

\title{\boldmath Emissivity measurements of CuCrZr alloy}

\author[a,1]{J. Song\note{Corresponding author.}}
\author[a]{N. Bultman}
\author[a]{M. Reaume}
\author[a]{S. I. Eom}
\author[a]{W. Franklin}
\author[a]{M. Patil}
\author[a]{\\R. Quispe-Abad}
\author[a]{E. Wakai}
\author[a]{M. Vargas Vallejo}

\affiliation[a]{Facility for Rare Isotope Beams, Michigan State University,\\ 640 South Shaw Lane,
  East Lansing, MI 48824, U.S.A}       
\emailAdd{songj@frib.msu.edu}

\abstract{
  The Facility for Rare Isotope Beams (FRIB)
  heavy-ion accelerator, in user operation since 2022, produces rare isotope beams via interactions of
  high-intensity stable ion beams with a graphite production target.
  Approximately 20–40$\%$ of the primary beam power is deposited
  in the target, while the remaining 60–80$\%$ is absorbed by the beam dump.
  The minichannel beam dump (MCBD), currently operated at 20 kW and designed for operation up to 50 kW,
  uses CuCrZr alloy absorber plates. Thermal validation and thermal cycling tests of the MCBD were conducted
  at the Applied Research Laboratory (ARL) at Pennsylvania State University.
  Temperature measurements were obtained from an infrared (IR) camera.
  Since accurate temperature determination requires reliable emissivity values,
  the emissivity of CuCrZr was measured using the IR camera validated against thermocouple reference temperatures
  up to approximately 650 $^{o}$C.
  The measurements were conducted under a vacuum level of approximately 10$^{-5}$ torr to minimize emissivity variations
  due to surface oxidation.  The emissivity of CuCrZr was determined to be 0.057 $\pm$ 0.009 using a constant fit to
  the measured data over the surface temperature range from 100--650 $^{\circ}\text{C}$. 
}
\begin{document}
\maketitle
\flushbottom
\section{Introduction}\label{sec:intro}
\begin{figure*}
  \begin{center}
    \begin{overpic}[width=0.88\textwidth]{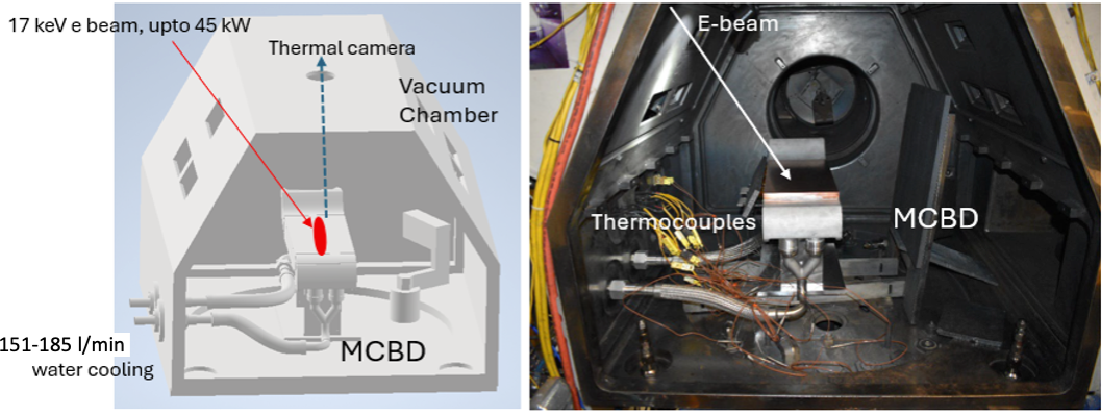}
      \put(43.8,33){\colorbox{white}{\small a)}}
      \put(96,33){\colorbox{white}{\small b)}}
    \end{overpic}
    \caption{
      a) 3D model of the MCBD thermal test set up showing the electron beam (e-beam) irradiation, IR camera position,
      and water-cooling configuration.
      b) Photo of the experimental setup inside the vacuum chamber, including the MCBD, thermocouples,
      and beam interaction region.
    }
    \label{fig:0}
  \end{center}
\end{figure*}

The FRIB, a world-leading rare isotope research facility, is advancing toward its design
beam power of 400 kW. Since the start of user operations in 2022, FRIB has delivered high-intensity stable ion beams ranging
from oxygen to uranium, with the operational beam power gradually increasing from 1 kW to 20 kW \cite{int_1,int_2}.
Further power upgrades are currently in progress.
During power ramp-up operation, the beam dump plays a critical role, as approximately 60–80$\%$ of the beam power
is deposited in this component. Up to 10 kW, a static aluminum beam dump \cite{int_3} was sufficient.
However, beyond this power level, an intermediate beam dump becomes necessary due to the limited thermal performance of aluminum.
To overcome this limitation, a minichannel beam dump (MCBD) \cite{int_4} was developed.
Compared with the previous design, the MCBD incorporates 2 mm-wide minichannels for water cooling, resulting in more than
a three times increase in the average convective heat transfer coefficient \cite{int_5}.
The design also employs a CuCrZr/Al2219 bimetal structure, where CuCrZr serves as the beam absorber and Al2219 aluminum alloy
forms the cooling channels.
Building on the authors’ prior experience with the neutron production target system for nuCARIBU \cite{int_6},
the design concept of this system was further developed for FRIB target and beam-dump R$\&$D activities,
based on established minichannel heat-sink principles \cite{int_7}.
Preliminary thermal validation results were reported in \cite{int_8},
where the experimental setup is shown in Fig. \ref{fig:0}.
Electron-beam (e-beam) testing was performed at the ARL at Pennsylvania State University.
Temperature measurements were carried out using an infrared (IR) camera.

Infrared thermography is a non-contact temperature measurement technique based on the detection of thermal radiation emitted
from a material surface. The spectral radiance of an ideal blackbody is described by Planck’s law,
\begin{equation*}
B_{\lambda}(T)=\frac{2hc^{2}}{\lambda^{5}}
\frac{1}{\exp\left(\frac{hc}{\lambda k_{\mathrm{B}}T}\right)-1},
\end{equation*}
where $T$, $h$, $c$ and $k_{B}$ are
the absolute temperature, Planck`s constant, the speed of light, and the Boltzmann constant, respectively.
By integrating the emitted radiation over wavelength and the hemispherical solid angle,
the radiative power emitted per unit area
from a surface can be expressed by the Stefan-Boltzmann relation,
\begin{equation}
p=\varepsilon \sigma T^{4},
\label{eq:1}
\end{equation}
where $p$, $\varepsilon$ and $\sigma$ are the radiative power emitted per unit area, the surface emissivity,
and the Stefan-Boltzmann constant, respectively.
The emissivity $\varepsilon$ represents the ratio of the radiation emitted by a real surface to that emitted by
an ideal blackbody at the same temperature.

Accurate beam dump temperature measurement is required to validate the thermal model and confirm that the MCBD operates
below the allowable temperature limit.
Because the allowable absorber temperature is constrained by material thermal performance,
a temperature uncertainty at the few-percent level is sufficient to distinguish meaningful deviations between
the measured temperature and thermal-model predictions, which typically agree within approximately 5$\%$
in the MCBD thermal validation tests.
Accurate emissivity determination is therefore essential for reliable IR-based temperature measurements and for evaluating
beam dump thermal performance. In particular, the low emissivity of CuCrZr and its possible temperature dependence can
introduce significant uncertainty in IR-based temperature measurements if not properly characterized. This work presents
the experimental determination of the emissivity of CuCrZr alloy over a wide temperature range up to approximately  650$^{\circ}\text{C}$. 
The results provide essential input parameters for accurate temperature evaluation and improve the reliability
of thermal validation of the MCBD under high-power beam operation.

\section{Apparatus}
The existing MCBD thermal test setup was adapted for emissivity measurements of the CuCrZr absorber used in the MCBD.
The experiments were performed inside a vacuum chamber, as shown in Fig. \ref{fig:0},
both to reproduce the vacuum operating environment of the MCBD and  to minimize surface oxidation
during high-temperature measurements.
Two independent heating systems were used. A laboratory hot plate was used for temperatures up to approximately 
300 $^{\circ}\text{C}$, while e-beam heating was used for higher temperatures up to approximately  650 $^{\circ}\text{C}$.
The temperature was measured using Type-K thermocouples and an IR camera,
as summarized in Table \ref{tab:1}.
The maximum temperature was limited by the operational range of the thermocouples.

\begin{table}[h]
  \centering
  \scriptsize
  \caption{Experimental samples and equipment specifications. Here NEDT denotes noise equivalent differential temperature.
  The hot plate is rated up to 500 $^{\circ}\text{C}$, but the maximum temperature reached in this work was approximately  310 $^{\circ}\text{C}$.}
  \label{tab:1}
  \begin{minipage}[t]{0.85\textwidth}
    \vspace{0pt}
    \centering
    \setlength{\tabcolsep}{6pt}
    \resizebox{\linewidth}{!}{
      \begin{tabular}{@{}l r@{\hspace{6pt}}|l r@{}}
        \hline
        \textbf{IR camera} & & \multicolumn{2}{l}{\textbf{Samples}} \\
        \hline
        Manufacturer & FLIR & Material & CuCrZr(C18150) \\
        Detector type & InSb & Dimensions & 50.8$\times$50.8 mm$^{2}$ \\
        \cline{3-4}
        Model & X6801sc & \multicolumn{2}{l}{\textbf{Hot plate}} \\
        \cline{3-4}
        Spectral range & 3.0--5.0 $\mu$m & Manufacturer & VWR \\
        Frame rate & 0.0015--520 Hz & Heating range & up to 500 $^{\circ}$C \\
        \cline{3-4}
        Temperature range & 10--2000 $^{\circ}$C & \multicolumn{2}{l}{\textbf{IR window}} \\
        \cline{3-4}
        Resolution & 640$\times$512 & Material & ZnSe\\
        Filtering & 4-position motorized filter wheel & Coating & uncoated\\
        NEDT & $<$ 20 mK & Thickness & 12.7 mm\\
        Accuracy & $\pm$ 1 $\%$ (typical) & Transmission & $\approx$ 70$\%$\\
        \hline
      \end{tabular}
    }
  \end{minipage}
\end{table}

\subsection*{Method}\label{method}
\begin{figure*}
  \begin{center}
    \begin{overpic}[width=0.88\textwidth]{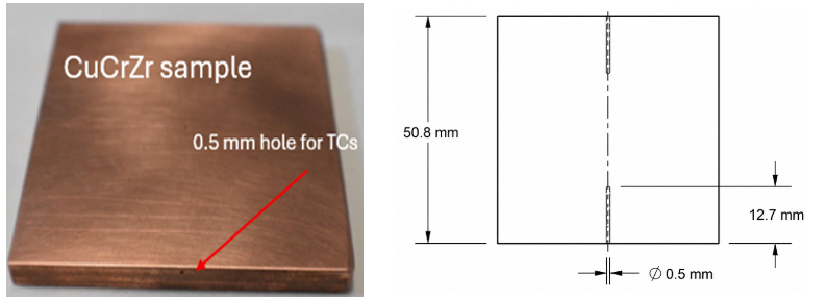}
      \put(0.2,38){\colorbox{white}{\small a)}}
      \put(58,38){\colorbox{white}{\small b)}}
    \end{overpic}
    \caption{
      a) Photo of the CuCrZr alloy sample showing the hole located near the surface for the 0.5 mm thermocouples.
      b) Technical drawing of the sample indicating the dimensions: 50.8$\times$50.8 mm$^{2}$ area, 6.35 mm thickness,
      a 0.6 mm diameter hole located close to the surface (1.3 mm distance), and a hole length of 12.7 mm.
    }
    \label{fig:01}
  \end{center}
\end{figure*}

The emissivity was determined by adjusting the emissivity setting of the IR camera until the measured temperature
matched the temperature measured by thermocouples positioned as close as possible to the sample surface.
To determine the appropriate emissivity value, the emissivity parameter was varied over the range of 0.04-0.07
to establish the temperature-emissivity dependence and identify the emissivity corresponding
to the thermocouple reference temperature.
Polished CuCrZr samples (see Fig. \ref{fig:01} and Sec. \ref{samplepre})
were prepared and placed inside the vacuum chamber to minimize surface oxidation and to reproduce the vacuum
operating environment of the MCBD during the high temperature measurements.
To verify the consistency of the emissivity measurements, two independent heating methods were employed.
The hot plate provided heating from the bottom surface, while the e-beam provided surface heating from the top.
These two heating configurations resulted in different temperature gradients between the surface
and thermocouple positions.
Despite these differences, the extracted emissivity values remained consistent within he required measurement precision.
For the present application, the required accuracy of the IR temperature evaluation was within 5$\%$,
corresponding to an emissivity difference of approximately  0.01.
The differences between the emissivity values obtained from the hot-plate and e-beam measurements were within this range.
To minimize temperature non-uniformity, emissivity values were extracted only from regions with relatively
uniform temperature distributions.
For the e-beam heating, a uniform beam profile was applied.
The average uncertainty in the IR temperature measurement was typically 1–2$\%$.
The vacuum pressure during the measurements was approximately  $5\times  10^{-5}$ torr.
Type-K thermocouples were positioned within 1.3 mm of the sample surface
to provide near-surface reference temperatures and minimize temperature gradient effects.
For the IR analysis, each region of interest was defined around the corresponding thermocouple location,
and the temperature distribution within each region was relatively uniform.
The typical uncertainty in the temperature measurements, including contributions from both the thermocouples and the IR camera, was 1–2$\%$.
The IR camera monitored the sample through a ZnSe IR window.
For each measurement, the samples were heated to the target temperature using either the hot plate or e-beam heating,
and data were recorded only after thermal equilibrium was reached, as confirmed by stable thermocouple readings.
The emissivity determination procedure was repeated at multiple temperatures with multiple samples
to ensure the reproducibility.

\section{Measurement}

\subsection*{Sample preparation}\label{samplepre}

CuCrZr samples with dimensions of 50.8 $\times$ 50.8 mm$^{2}$ and a thickness of 6.35 mm were prepared
for the emissivity measurements, as shown in Fig. \ref{fig:01}.
Two holes were machined for thermocouple installation, located within 1.3 mm of the top surface
and with a depth of 12.7 mm.
The sample surfaces were mechanically polished using a Scotch-Brite pad to ensure
consistent emissivity measurements over the surface area.
Since surface roughness strongly influences the emissivity, the surface finish was manually polished to
achieve reasonably consistent conditions. 
The average surface roughness (Ra) was approximately  0.7 $\mu$m.

\subsection*{IR Window Transmission Coefficient}\label{Measurementmethod}

\begin{figure}[h]
  \begin{center}
    \begin{overpic} [width=0.7\textwidth] {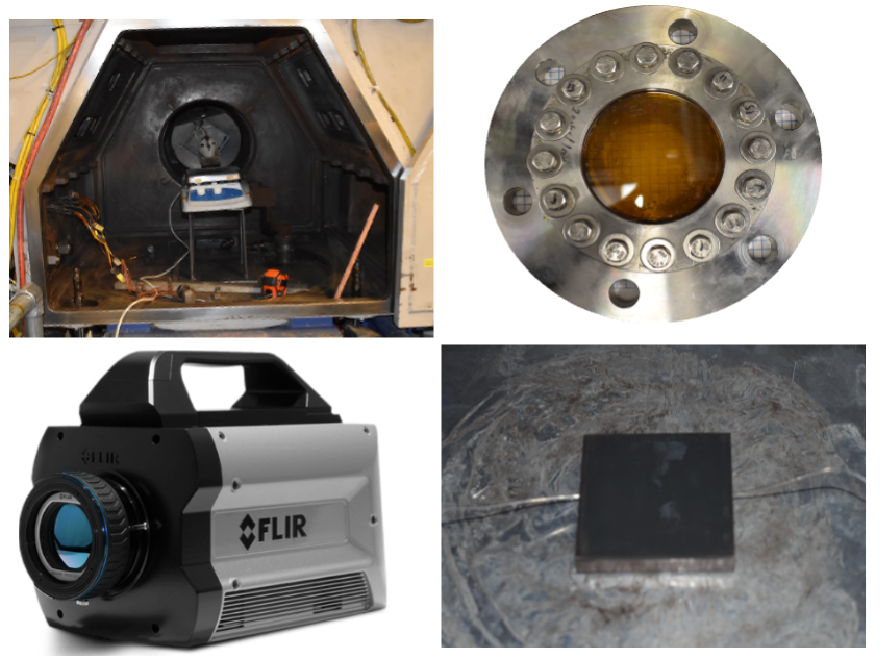}
      \put(1,68.5){\colorbox{white}{\small a)}}
      \put(49.3,68.5){\colorbox{white}{\small b)}}
      \put(1,32.5){\colorbox{white}{\small c)}}
      \put(49.3,32.5){\colorbox{white}{\small d)}}
    \end{overpic}
    \caption{
      Photos of the experimental setup.
      a) Vacuum chamber showing the hot plate and sample with two thermocouples installed.
      b) 12.7-mm-thick uncoated ZnSe IR window.
      c) FLIR IR camera.
      d) CuCrZr sample with a high-emissivity coating and two thermocouples mounted.
    }
    \label{fig:1}
  \end{center}
\end{figure}

To accurately determine the temperature using the IR camera, two parameters must be defined:
the transmission coefficient of the IR window and the emissivity of the sample in the IR camera setting.
Therefore, the transmission coefficient of the IR window was experimentally determined.
The experimental setup is shown in Fig. \ref{fig:1}.
The transmission measurement required temperature measurements both with and without the IR window.
Since vacuum conditions cannot be achieved without the IR window installed,
these measurements were performed in air.
To minimize emissivity variations caused by surface oxidation under these conditions, a CuCrZr sample
with a high-emissivity coating was used. The transmission coefficient was determined
by comparing the temperature measurements obtained with and without IR window.

\begin{figure*}
  \begin{center}
    \begin{overpic} [width=0.32\textwidth] {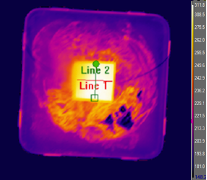}
    \end{overpic}
    \begin{overpic} [width=0.31\textwidth] {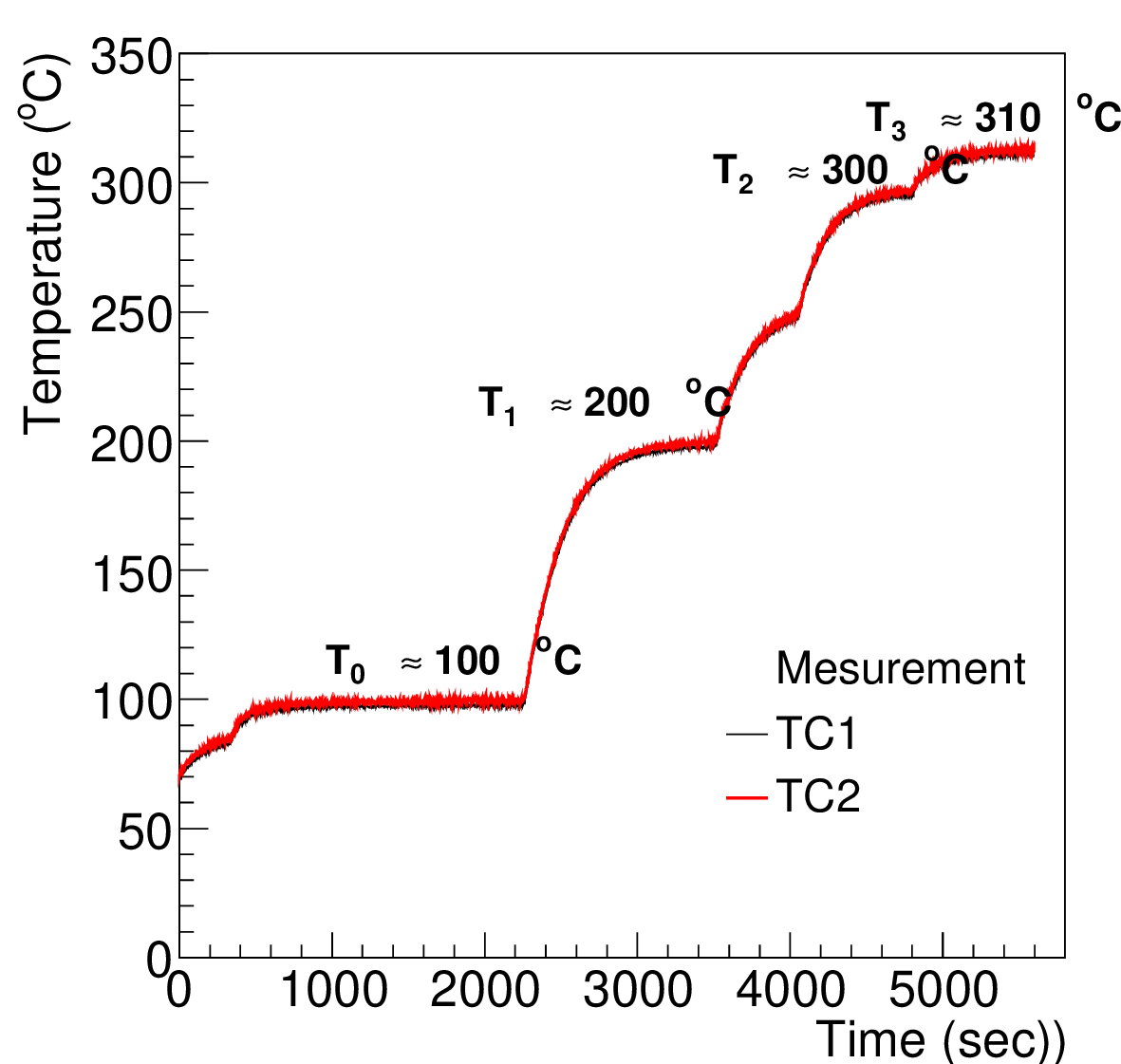}
    \end{overpic}
    \begin{overpic} [width=0.31\textwidth] {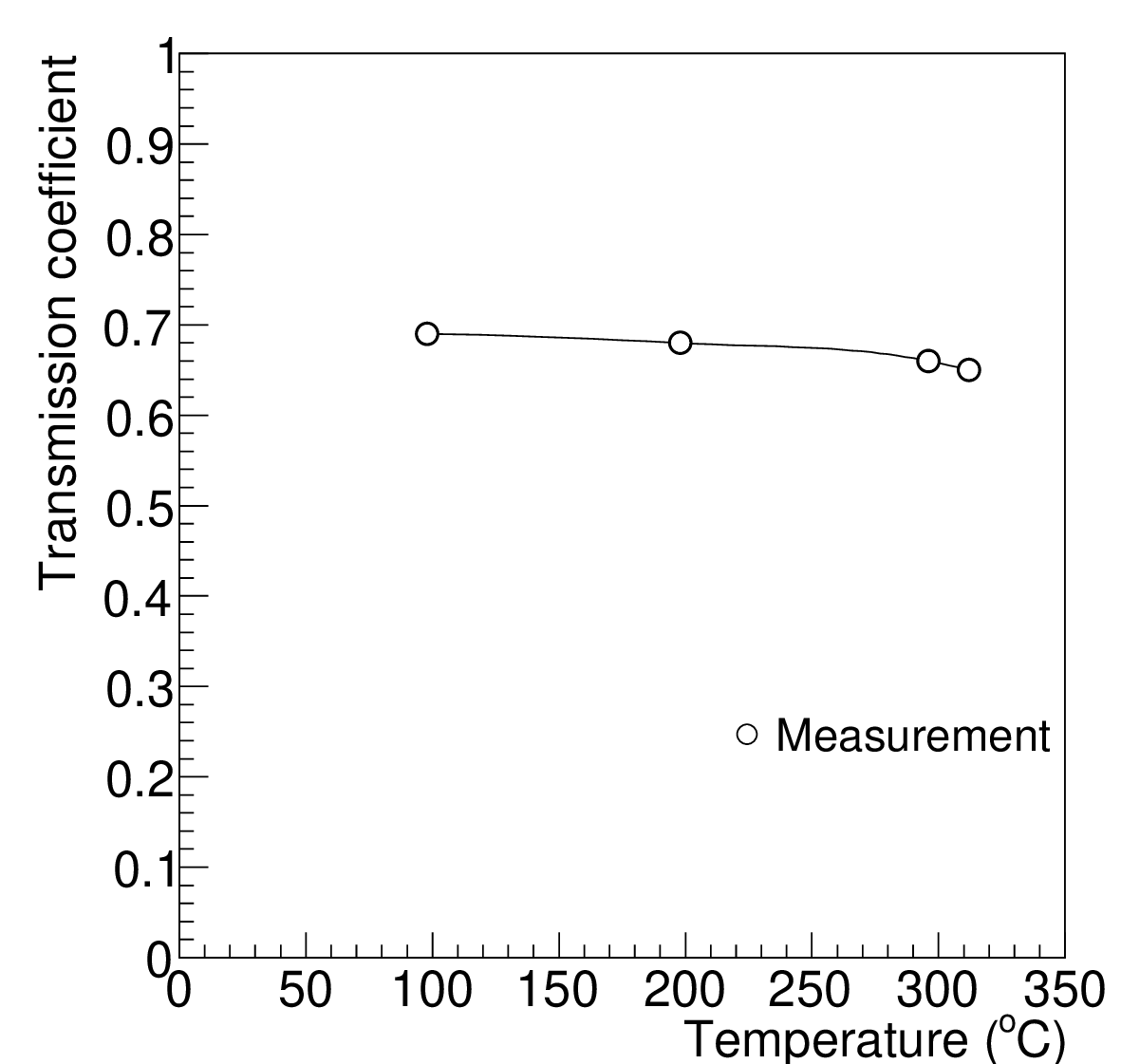}
      \put(-211,82){\colorbox{white}{\small a)}}
      \put(-85,82){\colorbox{white}{\small b)}}
      \put(17,82){\colorbox{white}{\small c)}}
    \end{overpic}
    \caption{
      a) Thermal image obtained using an IR camera at approximately  300 $^{\circ}\text{C}$.
      b) Measured temperature evolution as a function of time obtained from thermocouples (TC1 and TC2).
      c) Measured IR window transmission coefficient as a function of temperature up to 310 $^{\circ}\text{C}$.
    }
    \label{fig:2}
  \end{center}
\end{figure*}
The transmission coefficients of the IR window were measured at four different temperatures:
100, 200, 300, and 310 $^{\circ}\text{C}$.
The temperature of 310 $^{\circ}\text{C}$ represents the upper limit achievable with the laboratory hot plate
used in	this study.
An example of the thermal image	obtained using the IR camera is shown in Fig. \ref{fig:2} a).
Temperature data were selected for analysis after reaching thermal steady-state conditions,
as confirmed by	stable	thermocouple readings (see Fig. \ref{fig:2} b).
The transmission coefficient was determined to be 0.68 by comparing temperatures measured with and without
the IR window at each temperature point, as shown in Fig. \ref{fig:2} c). This procedure enabled assessment	
of the temperature dependence of the IR	window transmission coefficient.
The transmission coefficient value of 0.68 was adopted as a fixed parameter in the emissivity measurements.
The IR-window transmission coefficient showed a slight temperature dependence, as shown in Fig. \ref{fig:2} c),
and can also depend on wavelength. In the present work, a value of 0.68 was adopted as an effective parameter
for the specific measurement configuration used here.
Therefore, this value should not be interpreted as a temperature-independent material constant,
but as a practical approximation for the present IR analysis.
Since it was measured only up to 310 $^{\circ}\text{C}$,
possible additional temperature dependence at higher temperatures was not explicitly characterized in the present work.
\subsection*{Emissivity of CuCrZr alloy}\label{em}

\begin{figure}[b]
  \begin{center}
    \begin{overpic} [width=0.98\textwidth] {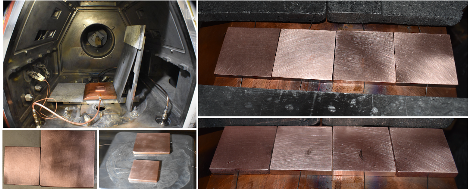}
      \put(0,38.5){\colorbox{white}{\small a)}}
      \put(0,11){\colorbox{white}{\small b)}}
      \put(21,11){\colorbox{white}{\small c)}}
      \put(42,38.5){\colorbox{white}{\small d)}}
      \put(42,13.5){\colorbox{white}{\small e)}}
    \end{overpic}
    \caption{
      a) Emissivity measurement setup showing the sample placed in thermal contact
      with the water-cooled copper beam dump.
      b)--c) Two samples before and after hot-plate testing
      d)--e) Four samples before and after e-beam testing.
    }
    \label{fig:4}
  \end{center}
\end{figure}
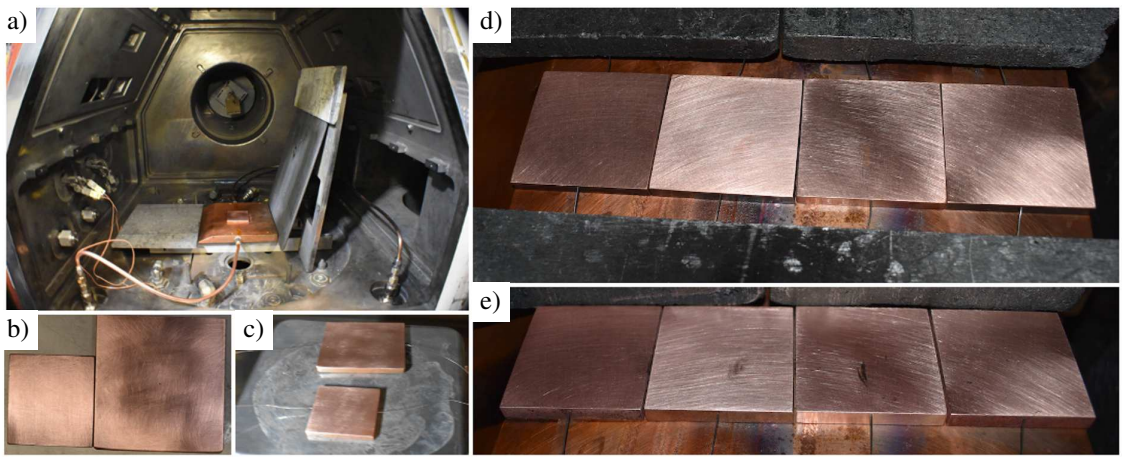

\begin{figure}
  \begin{center}
    \begin{overpic} [width=0.9\textwidth] {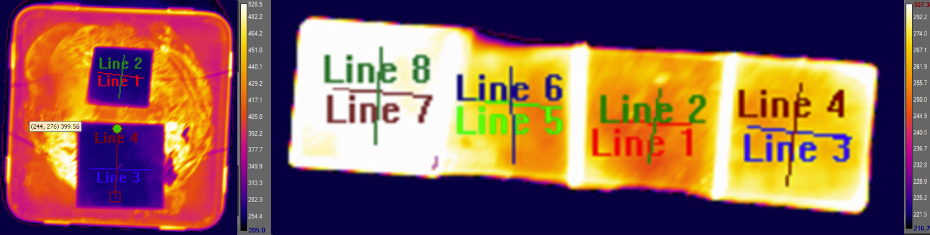}
    \end{overpic}
    \begin{overpic} [width=0.325\textwidth] {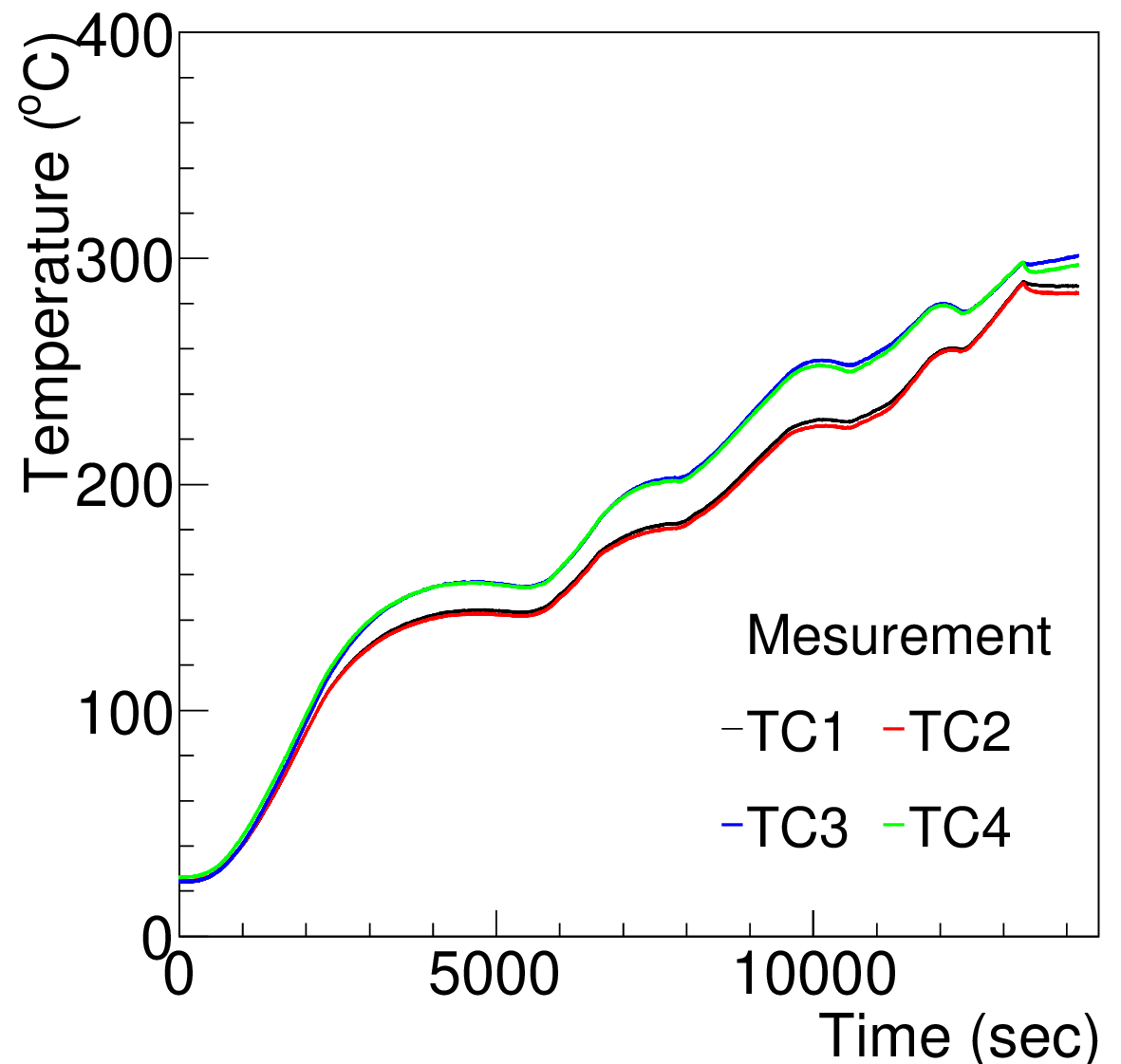}
    \end{overpic}
    \begin{overpic} [width=0.325\textwidth] {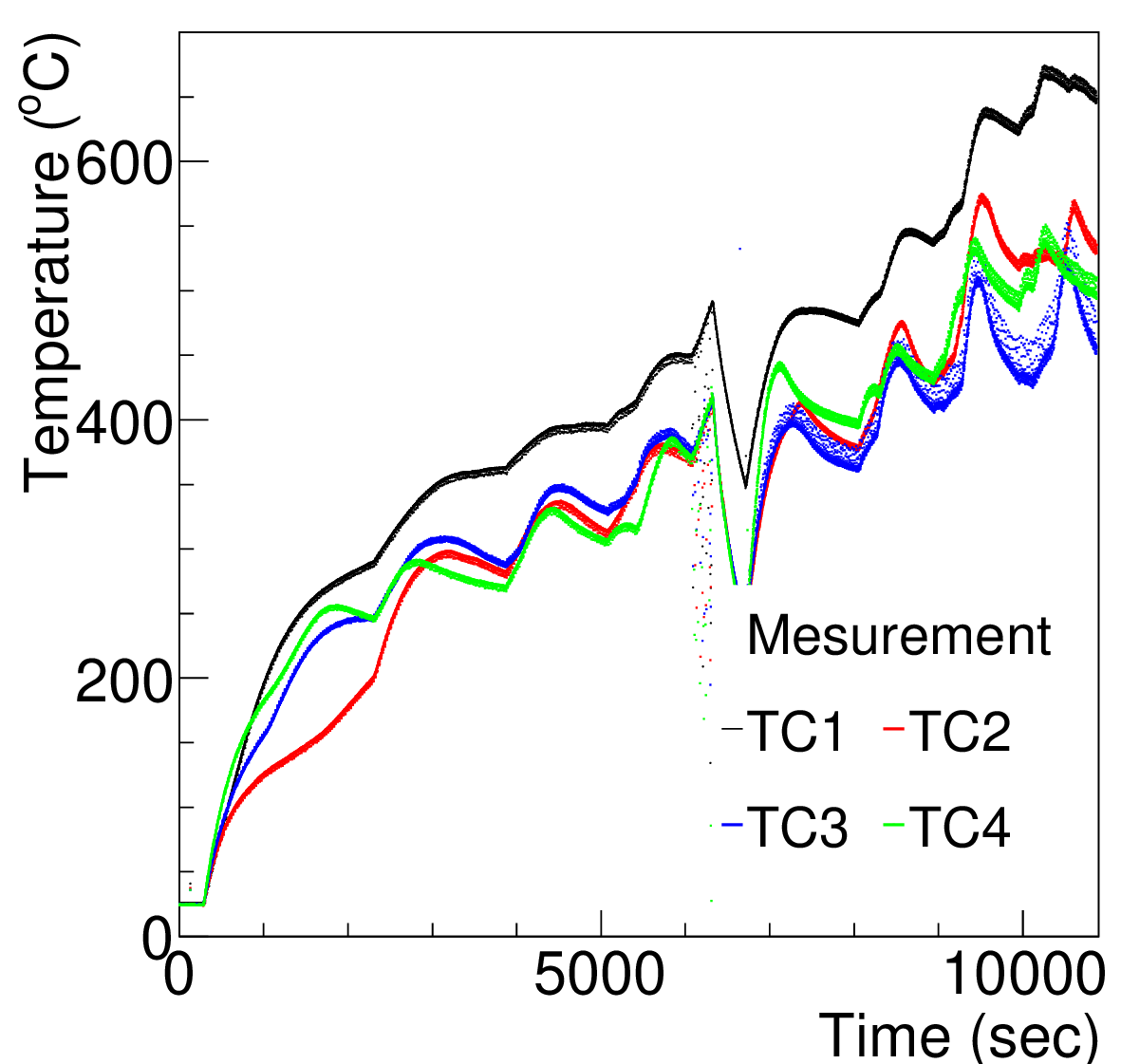}
    \end{overpic}
    \begin{overpic} [width=0.325\textwidth] {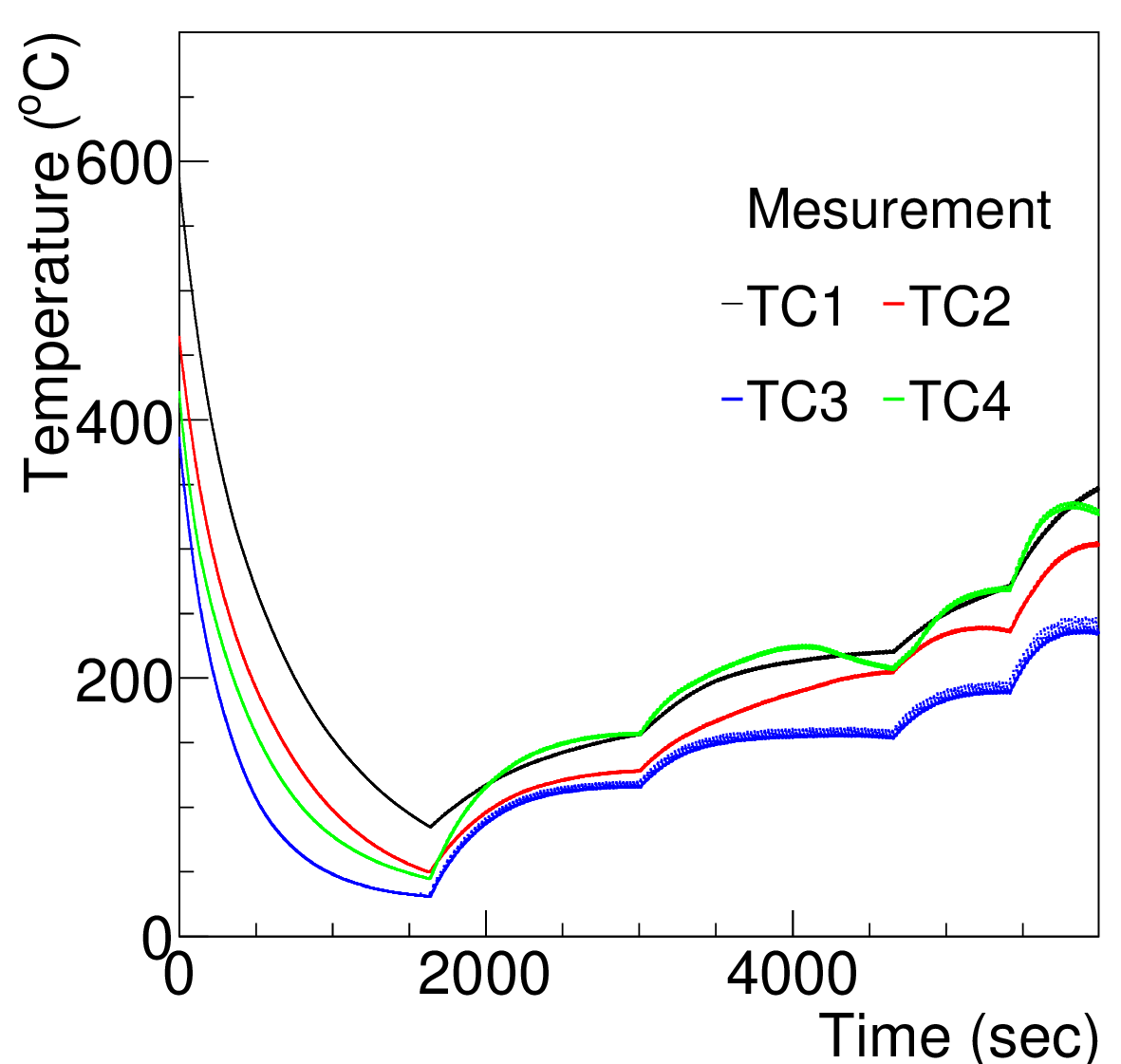}
      \put(-193,158){\colorbox{white}{\small a)}}
      \put(-111,158){\colorbox{white}{\small b)}}
      \put(-188,82){\colorbox{white}{\small c)}}
      \put(-85,82){\colorbox{white}{\small d)}}
      \put(20,82){\colorbox{white}{\small e)}}
    \end{overpic}
    \caption{
      Thermal images of the samples obtained using an IR camera for hot-plate heating (a) and e-beam heating (b).
      Measured temperature evolution as a function of time obtained from thermocouples
      during hot-plate heating (c), e-beam heating (d), and e-beam reheating (e) after cooling under vacuum.
    }
    \label{fig:5}
  \end{center}
\end{figure}

Two independent heating methods were employed to perform the emissivity measurements, as described in Sec. \ref{method}.
Hot-plate heating was used for temperatures up to 300 °C, while e-beam heating was used
for higher temperatures up to approximately  650 $^{\circ}\text{C}$. The corresponding experimental setups are illustrated
in Figs. \ref{fig:1} a) and Fig. \ref{fig:4} a).
For hot-plate heating, two samples of different sizes (50.8$\times$50.8 mm$^{2}$ and 63.5$\times$63.5 mm$^{2}$)
were placed near the center of the hot plate to minimize radial temperature gradients, as shown in Fig. \ref{fig:5} a).
Type-K thermocouples were positioned near the sample surfaces (within 1.3 mm of the surface) to provide
near-surface reference temperatures for comparison with the IR camera measurements.
The temperature variation across the samples was typically within 1–2$\%$.
Measurements were conducted at four steady-state temperatures (150, 200, 250, and 300 $^{\circ}\text{C}$).
At each temperature step, data were recorded only after thermal equilibrium was reached,
as verified by stable thermocouple readings over time (Fig. \ref{fig:5} c)).
For e-beam heating, four samples were placed in thermal contact with the copper beam dump, as shown in Fig. \ref{fig:4} d).
Graphite blocks were used to shield the thermocouples from direct beam exposure and to reduce electrical interference.
Although a uniform beam profile was applied, small temperature variations among the samples were observed due to
spatial differences in heat removal caused by the internal water-cooling channels of the copper beam dump.
The temperature variation across the samples was also typically within 1–2$\%$.
In both heating configurations, the emissivity was determined by adjusting the IR camera emissivity setting,
following the same procedure described in Sec. \ref{method}.

\section{Results}\label{Results}
\begin{figure*}
  \begin{center}
    \begin{overpic} [width=0.94\textwidth] {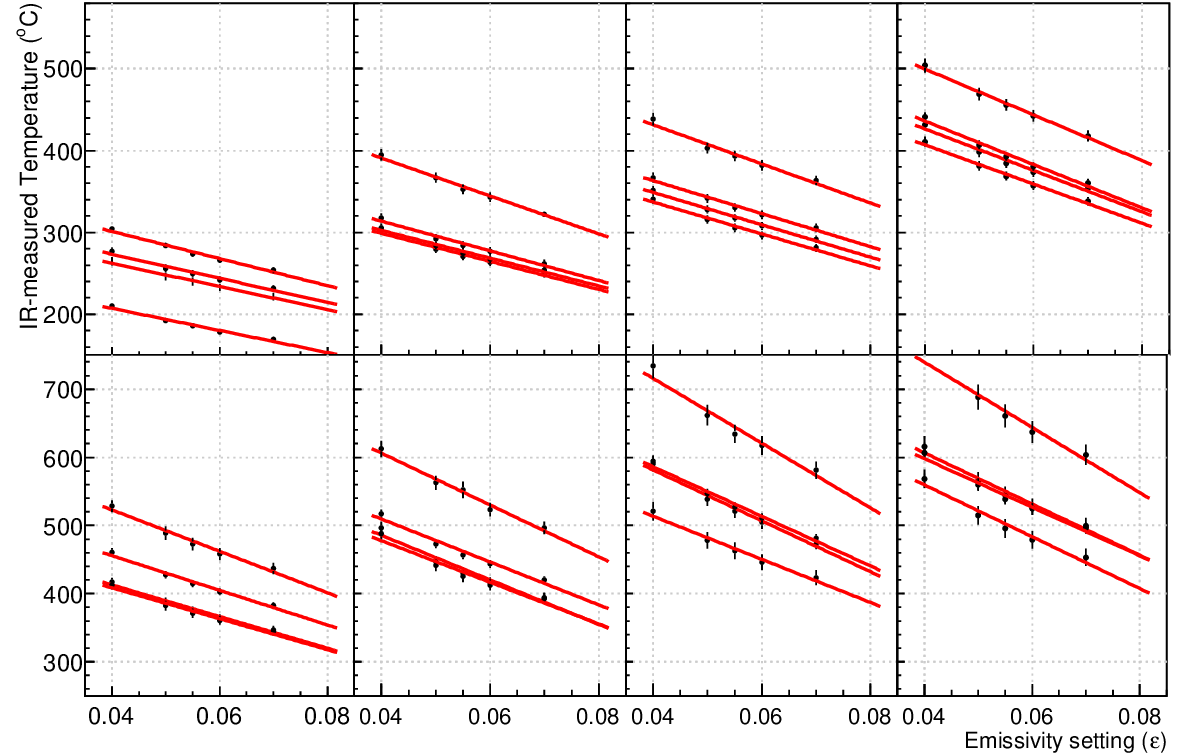}
    \end{overpic}
    \begin{overpic} [width=0.94\textwidth] {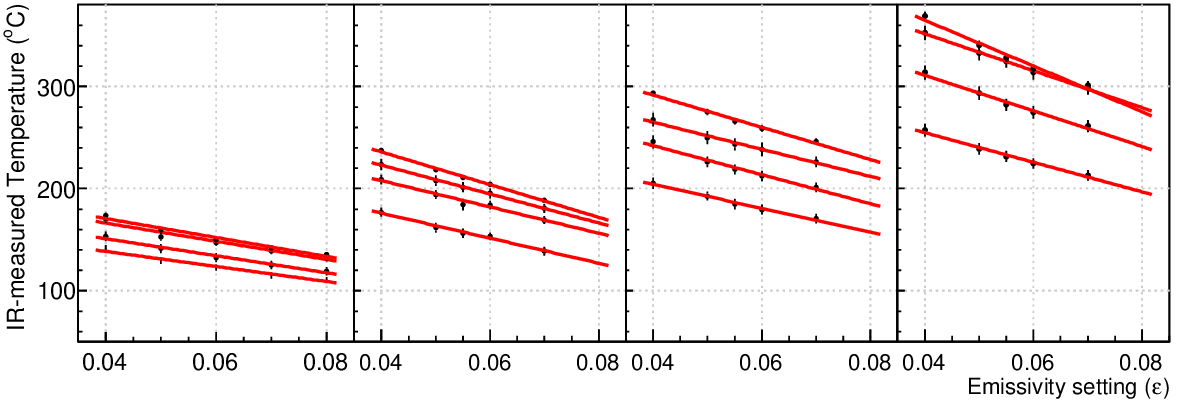}
    \end{overpic}
    \begin{overpic} [width=0.94\textwidth] {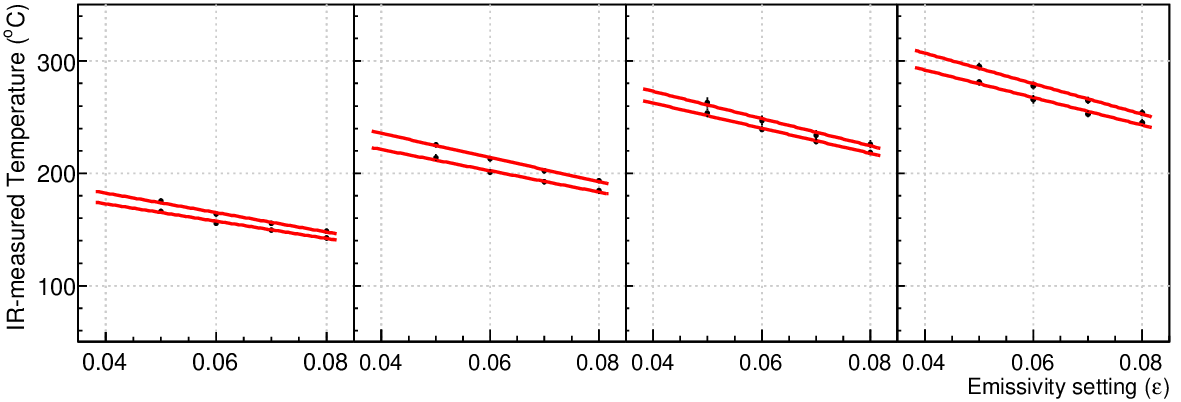}
      \put(8,128){\colorbox{white}{\small a)}}
      \put(7,64){\colorbox{white}{\small b)}}
      \put(7,30){\colorbox{white}{\small c)}}
    \end{overpic}
    \caption{IR-measured temperatures as a function of the emissivity setting of the IR camera.
      The emissivity setting was scanned from 0.04 to 0.07 or 0.05 to 0.08, depending on the measurement condition. 
      Black markers represent IR-measured temperatures, with vertical error bars indicating temperature uncertainties
      obtained using the IR camera.
      Red lines show linear fits describing the relationship between IR-measured temperature and emissivity setting.
      a) e-beam heating up to approximately  650 $^{\circ}\text{C}$, 
      b) e-beam reheating after cooling up to approximately  350 $^{\circ}\text{C}$,
      and c) hot-plate heating up to approximately 300 $^{\circ}\text{C}$.
    }
    \label{fig:6}
  \end{center}
\end{figure*}

\begin{figure*}
  \begin{center}
    \begin{overpic} [width=0.98\textwidth] {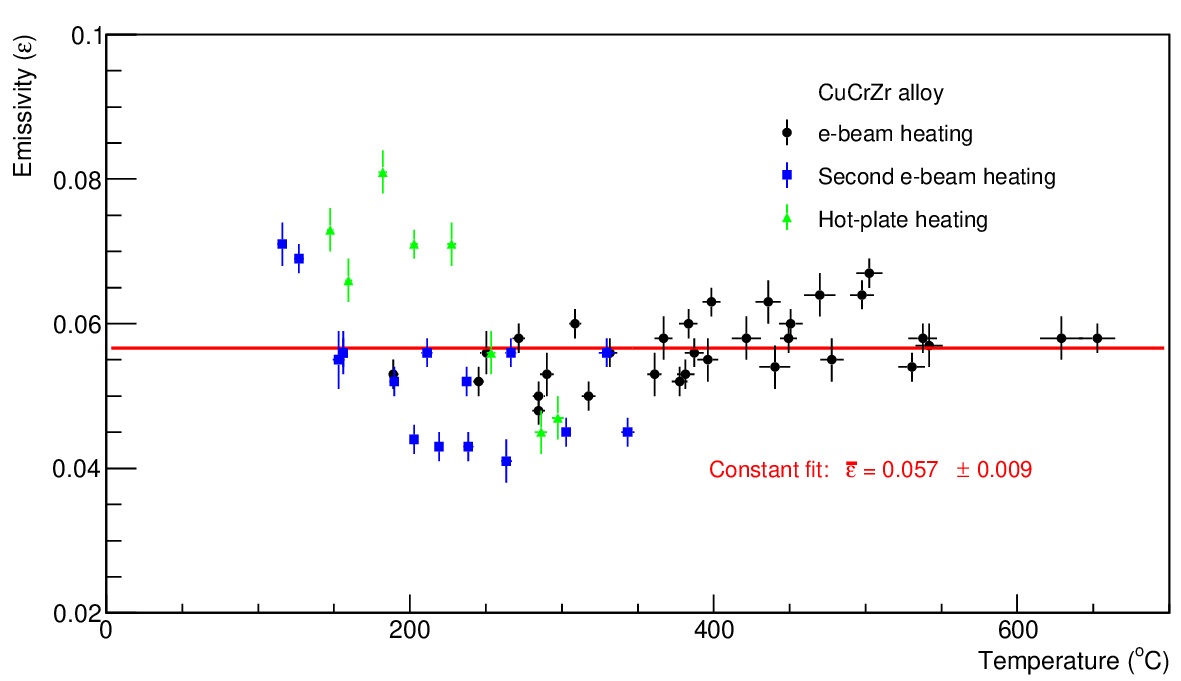}
    \end{overpic}
    \caption{Measured emissivity of the CuCrZr alloy as a function of surface temperature.
      The vertical error bars represent the measurement uncertainties in emissivity at each temperature,
      while the horizontal bars indicate the temperature uncertainties.
      The black, blue, and green markers indicate measurements from the
      first e-beam heating, the second e-beam heating following cooling, and the hot plate heating, respectively.
      The average emissivity over the this temperature range was determined from a constant fit to the data.
    }
    \label{fig:7}
  \end{center}
\end{figure*}

The individual emissivities of the CuCrZr samples were determined over the temperature range of 100 to 650 $^{\circ}\text{C}$
using both hot-plate and e-beam heating. The emissivity values were confirmed
by adjusting the IR camera emissivity setting until the measured temperatures matched
the reference temperatures obtained from thermocouples. Two and four CuCrZr samples were used for the hot-plate heating
and e-beam heating measurements, respectively.
A total of 56 data points were obtained as shown in Fig. \ref{fig:6}: 32 from the first heating cycle (200-650 $^{\circ}\text{C}$),
16 from the second heating cycle (100-350 $^{\circ}\text{C}$),
and 8 from hot plate heating (150-300 $^{\circ}\text{C}$).
IR camera temperatures were obtained by scanning the emissivity setting from 0.04 to 0.07 or 0.05 to 0.08,
depending on the measurement condition.
Because the detected radiance depends on both surface temperature and emissivity,
changing the emissivity setting in the IR camera changes the inferred IR temperature for the same measured signal.
Although the temperature-emissivity relation is not strictly linear, it was approximated locally by a
linear function over the narrow emissivity range scanned in this work.
Using the simplified relation $T(\varepsilon) = C\varepsilon^{-1/4}$ from the Stefan–Boltzmann relation
Eq. \ref{eq:1}, the first-order Talylor expansion around $\varepsilon_{o}$ gives,
\begin{equation}
T(\varepsilon) \approx T_{0} - \dfrac{T_{0}}{4\varepsilon_{0}}(\varepsilon - \varepsilon_{0}).
\end{equation}
Therefore, linear fits were used as a practical local interpolation to determine the emissivity corresponding to the thermocouple reference
temperature (see Fig. \ref{fig:6}).
The emissivity for each measurement was determined by matching the thermocouple temperature
to the corresponding linear temperature–emissivity relation.
The uncertainties associated with the IR camera and thermocouple temperature measurements were propagated to
estimate the emissivity uncertainty.
In this analysis, only temperature measurement uncertainties were included.
The emissivity uncertainty was calculated as,
\begin{equation}
\Delta \varepsilon = \dfrac{\Delta T}{\partial T / \partial \varepsilon},
\end{equation}
where $\Delta T$ = $\sqrt{(\Delta T_{IR})^{2}+(\Delta T_{TC})^{2}}$) is the combined one-standard-deviation
uncertainty of the IR camera and thermocouple temperature measurements.
A total of 56 emissivity measurements are shown in Fig. \ref{fig:7}.
The vertical and horizontal error bars represent emissivity and temperature uncertainties, respectively.
The black, blue, and green markers indicate measurements from the first e-beam
heating, the second e-beam heating following cooling, and the hot-plate heating.
The effective emissivity was obtained from a weighted constant fit using the individual measurement uncertainties.
No statistically significant monotonic temperature dependence was observed over the full temperature range
when the scatter among samples and heating cycles was included. The observed variation is therefore treated
as part of the effective uncertainty rather than fitted with a temperature-dependent model.
Thus, the constant-fit emissivity is interpreted as an effective engineering parameter for IR camera-based thermal measurements,
rather than as a fundamental material properties.
To calculate the emissivity uncertainty, the reduced $\chi^{2}$ was first determined as,
\begin{equation}
  \chi^{2}_{\nu} = \dfrac{\chi^{2}}{N-1} = \dfrac{1}{N-1}\sum_{i}^{N}\dfrac{(x_{i}-\varepsilon)^{2}}{\sigma_{i}^{2}}, 
\end{equation}
where $N$ is the number of measurements, and  $\sigma_{i}$ is the uncertainty of each emissivity measurement.
Since the individual measurement uncertainties were similar, a representative measurement uncertainty was
defined as,
\begin{equation*}
\sigma_{\mathrm{rep}} = \sqrt{\frac{1}{N-1}\sum_{i=1}^{N}\sigma_{i}^{2}}.
\end{equation*}
The effective emissivity uncertainty was obtained by scaling the representative uncertainty according to
the  $\chi_{\nu}^{2}$,
\begin{equation}
  \sigma_{\varepsilon} = \sigma_{\mathrm{rep}}\sqrt{\chi^{2}_{\nu}}.
  \label{eq:4-1}
\end{equation}
From this analysis, the emissivity of CuCrZr was determined to be
$\varepsilon = 0.057\pm0.009$ over the studied temperature range of 100--650 $^{\circ}\text{C}$.

\section{Conclusion}
\label{cd}
The emissivity of polished CuCrZr alloy was experimentally characterized to improve the accuracy of
IR-based temperature measurements for beam dump thermal validation \cite{int_8}.
The transmission coefficient of the IR window was measured, and a fixed value of 0.68 was adopted for the emissivity measurements.
Since the IR window transmission coefficient was measured only up to 310 $^{\circ}\text{C}$
and may depend on wavelength, the adopted constant value should be regarded as an effective parameter
for the present IR camera configuration.
A total of 56 emissivity data points, together with their associated uncertainties, were obtained by comparing thermocouple
and IR camera temperatures using two heating methods: laboratory hot-plate heating and e-beam heating.
The measurements were conducted under a vacuum level of approximately  10$^{-5}$ torr
to minimize surface oxidation; the effect of lower vacuum levels
on surface oxidation is discussed in the Appendix.
The effective emissivity was determined from a weighted constant fit, and its uncertainty was estimated using Eq. \ref{eq:4-1}
with the reduced $\chi^{2}$.
The reduced $\chi^{2}$ was 11.56, indicating that the observed data scatter exceeds the estimated measurement uncertainties
likely due to systematic variation associated with surface condition difference among samples.
The resulting emissivity provides a practical input parameter for reliable IR-based temperature measurements of
CuCrZr beam dump surfaces for thermal validation. The measured value is slightly higher than typical literature values
for pure copper \cite{int_9}, which may be attributed to differences in material composition and surface condition.
\appendix
\section*{Appendix}
\label{appendix:1}
\begin{figure*}[h]
  \begin{center}
    \begin{overpic} [width=0.88\textwidth] {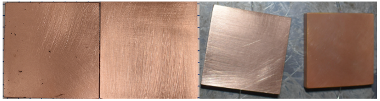}
      \put(0,24){\colorbox{white}{\small a)}}
      \put(52,24){\colorbox{white}{\small b)}}
    \end{overpic}
    \caption{
      Photos of the two samples before heating a) and after heating using the laboratory hot-plate
      under vacuum leak conditions of approximately  10$^{-1}$ tor b).,
    }
    \label{fig:8}
  \end{center}
\end{figure*}

\begin{figure*}[h]
  \begin{center}
    \begin{overpic} [width=0.68\textwidth] {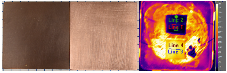}
      \put(0,30){\colorbox{white}{\small a)}}
      \put(32,30){\colorbox{white}{\small b)}}
      \put(64,30){\colorbox{white}{\small c)}}
    \end{overpic}
    \caption{
      Photos of the oxidized a) and polished b) samples.
     c)  An example of the IR temperature measurement at approximately  200 $^{\circ}\text{C}$ with an emissivity setting of 0.009. 
    }
    \label{fig:9}
  \end{center}
\end{figure*}

During the emissivity measurement for two samples, shown in Fig. \ref{fig:8},
a vacuum  leak occurred during the hot-plate testing.
Visible surface discoloration was observed.
After vacuum level was restored to approximately  10$^{-1}$ torr, the sample surfaces were repolished.
To quantify the effect of surface oxidation on emissivity, oxidized and polished samples, as shown in Fig \ref{fig:9} a) and b),
were compared.
A clear emissivity difference can be inferred from the temperature difference observed in the IR image
shown in Fig. \ref{fig:9} c).
The upper region corresponds to the polished sample, while the lower region corresponds to the oxidized sample.
The emissivities of the polished and oxidized samples were 0.05-0.08 and 0.3, respectively,
over the temperature range of 200-300 $^{\circ}\text{C}$.
\section*{ACKNOWLEDGMENTS}
This material is based upon work supported by the U.S. Department of Energy, Office of Science, Office of Nuclear
Physics and used resources of the Facility for Rare Isotope Beams (FRIB) Operations, which is a DOE Office
of Science User Facility under Award Number DE-SC0023633.

\end{document}